\documentclass[11pt]{article}

\usepackage[preprint]{acl}
\usepackage{times}
\usepackage{latexsym}

\usepackage[T1]{fontenc}
\usepackage[utf8]{inputenc}
\usepackage{multirow}
\usepackage{microtype}
\usepackage{inconsolata}
\usepackage{graphicx}
\usepackage{algorithm}
\usepackage{algorithmic}
\usepackage{amsmath}
\usepackage{amsfonts}
\usepackage{enumitem}
\usepackage{makecell}

\title{A Case-Driven Multi-Agent Framework for E-Commerce Search Relevance}

\author{Global E-Commerce Search Relevance Team \thanks{Corresponding Author, E-mail:  \texttt{wangfeiyi@bytedance.com}} \\
  ByteDance \\
}

\newcommand{\safeincludegraphics}[2][]{%
	\IfFileExists{#2}{%
		\includegraphics[#1]{#2}%
	}{%
		\fbox{\parbox[c][0.35\textheight][c]{\linewidth}{\centering Missing: \texttt{\path{#2}}}}%
	}%
}

\begin{document}
% \nocite{*}
\maketitle

\begin{abstract}
Relevance is a foundation of user experience in e-commerce search. We view relevance optimization as a closed-loop ecosystem involving multiple human roles: users who provide feedback, product managers who define standards, annotators who label data, algorithm engineers who optimize models, and evaluators who assess performance. Because improving relevance in practice means systematically resolving user-perceived bad cases, we ask a system-level question: can this ecosystem be reimagined by replacing its human roles with autonomous agents?
To answer this question, we propose a case-driven multi-agent framework that automates the pipeline from bad-case identification to resolution. The framework instantiates an Annotator Agent for multi-turn annotation, an Optimizer Agent for autonomous bad-case analysis and resolution, and a User Agent that identifies bad cases through conversational interaction, together forming an autonomous and continually evolving system. To make the framework practical in production, we further adopt a harness-engineering paradigm and build a unified retrieval-and-ranking relevance model for efficient training, an instruction-following relevance model for real-time case resolution, Global Memory to reduce information asymmetry across agents, a Deep Search Agent to target underestimation failures, and an agent-based chatbot for human--agent collaboration. Extensive human evaluation shows that the framework performs relevance-related tasks effectively, improves annotation accuracy, and enables more timely and generalizable bad-case resolution, indicating a practical paradigm for industrial search relevance optimization.
\end{abstract}

\section{Introduction}
\label{sec:introduction}

Relevance is a foundational requirement in e-commerce search because it sets the minimum quality bar for user experience.
At its core, the task is to decide whether a product (document) is relevant to a user query so that highly relevant products can be prioritized for display.

Existing research has mainly focused on improving the relevance model itself.
Earlier approaches typically rely on BERT-based models trained on large-scale human-annotated data \citep{devlin2019bert, nogueira2019passage},
while recent work explores more powerful large language models (LLMs) via supervised fine-tuning (SFT) and reinforcement learning with verifiable reward (RLVR) \citep{chung2024flan, ouyang2022instruct},
for either distillation or direct online serving (e.g., TaoSR1, ADORE) \citep{dong2026taosr1thinkingmodelecommerce, fang2025adoreautonomousdomainorientedrelevance}.
Although scaling model capacity is an effective modeling direction for increasingly complex cases, the practical iteration of search relevance in industry is fundamentally \emph{case-driven}: optimization starts from user-perceived bad cases and ends with their resolution.

From this perspective, relevance optimization is better viewed as a closed-loop ecosystem involving multiple human roles:
\textbf{users} report bad cases based on their search experience;
\textbf{annotators} label samples under predefined standards;
\textbf{product managers and evaluators} define, refine, and assess relevance criteria;
and \textbf{algorithm engineers} analyze cases and optimize models accordingly.
The conventional workflow proceeds as follows:
(1) a user reports a bad case;
(2) a product manager collects cases and updates standards;
(3) annotators label new samples;
(4) an algorithm engineer analyzes the case and improves the model;
(5) an evaluator assesses the updated model;
and (6) the model is deployed to resolve the case.

However, every role in this loop is inherently imperfect.
User feedback can be subjective, standards may fail to cover all scenarios, annotations may contain errors, and model updates can solve one issue while introducing another.
This motivates a broader systems question:
\emph{can the relevance ecosystem be reimagined by replacing these human roles with autonomous agents?}

To answer this question, we propose a case-driven multi-agent framework, depicted in Figure~\ref{fig:framework}, which comprises the following core elements:

\begin{figure*}[t]
  \centering
  \includegraphics[width=\textwidth]{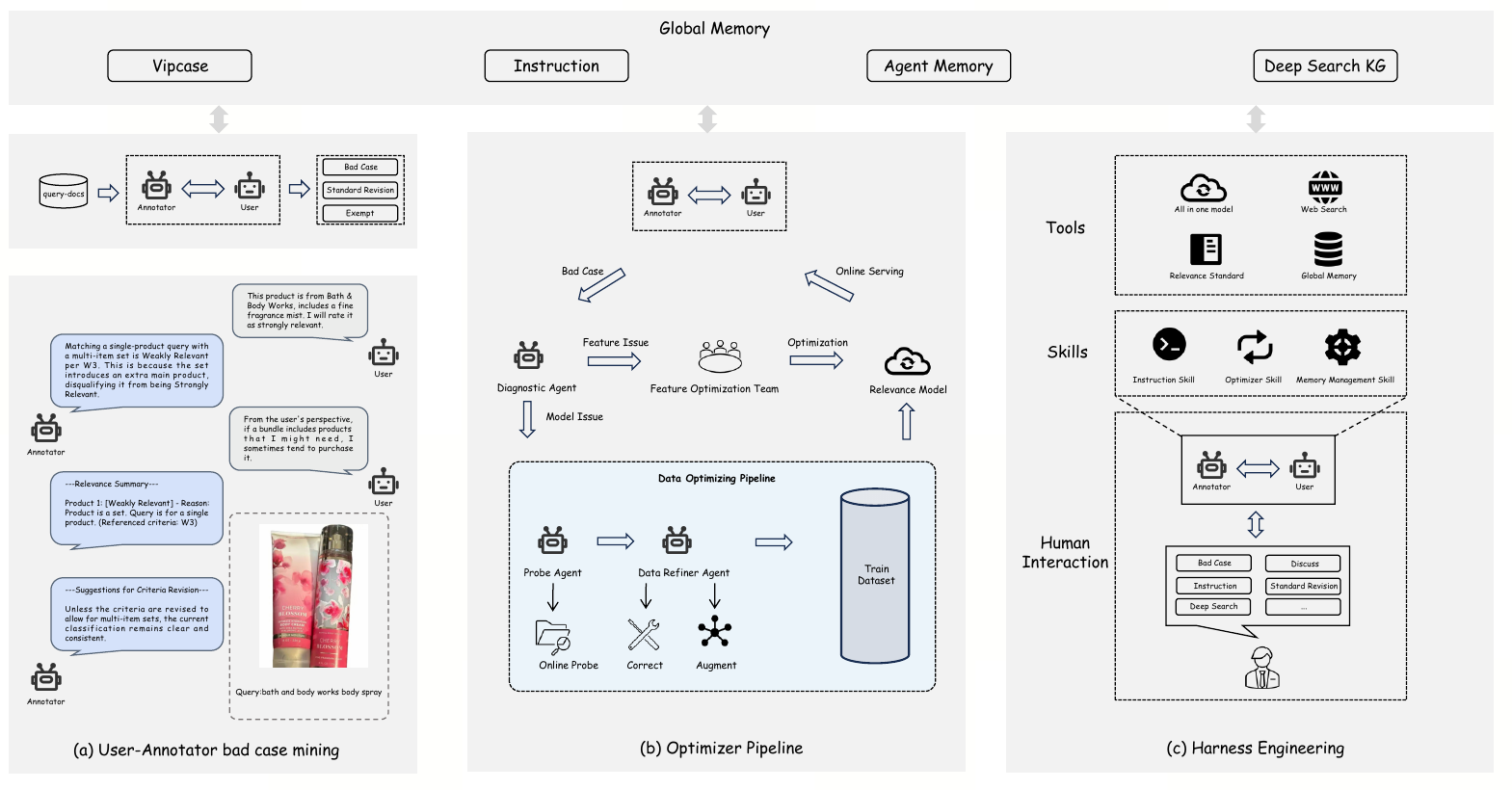}
  \caption{\textbf{Case-driven multi-agent framework for e-commerce search relevance.} The User Agent discovers high-value bad cases through conversational interaction; the Annotator Agent produces standard-grounded labels under evolving standards $S$ (and optional directives $I$, more in Appendix~\ref{sec:appendix_task_definition}); and the Optimizer Agent diagnoses failures and triggers data-centric repairs and retraining. An Automated Iteration Pipeline closes the loop by continuously sampling traffic, updating supervision, and deploying validated model updates, while Global Memory synchronizes standards and case status across agents.}
  \label{fig:framework}
\end{figure*}

The framework defines an end-to-end, auditable loop from bad-case discovery to diagnosis, repair, and deployment, enabling relevance to evolve through continual agent coordination rather than sporadic manual iteration.
\begin{enumerate}
    \item \textbf{An Automated Annotation Agent (Annotator):} an in-house, LLM-based annotator that achieves more precise sample annotation through multi-path reasoning and a generative relevance model (GRM).
    \item \textbf{A Model Optimization Agent (Optimizer):} an agentic, data-centric system that automates case analysis and resolution by transforming the conventional sample correction and augmentation workflow into an agent-driven process.
    \item \textbf{A Bad-Case Mining Agent (User):} a simulated consumer that reports cases and engages in dialogue with the Annotator Agent to provide feedback and suggest standard revisions.
    \item \textbf{An Automated Iteration Pipeline:} an integrated system that connects the User, Annotator, and Optimizer agents, enabling the self-evolution of relevance through automated discovery and resolution of bad cases.
\end{enumerate}

Motivated by harness engineering, we further build an efficient and reliable operating environment for these agents:
\begin{enumerate}
    \item \textbf{An All-In-One Relevance Model:} a unified model that integrates the retrieval, coarse ranking, and fine ranking stages, substantially reducing model training time.
    \item \textbf{Instruction-Following Capability:} an instruction-following relevance model that accepts real-time natural-language directives, allowing agents to resolve cases without retraining.
    \item \textbf{Global Memory:} a shared memory that breaks down communication barriers and information asymmetry among agents, enabling real-time awareness of standard updates and case resolution status.
    \item \textbf{Deep Search:} a Deep Search Agent that targets underestimation and recall failures in which truly relevant products are absent when the system encounters complex queries that rely on external knowledge rather than text matching.
    \item \textbf{Agent-Based Operations Chatbot:} a chatbot that exposes the agents in a human-facing interface so they can participate directly in daily work.
\end{enumerate}

Our framework has been deployed in production and has become a major driver of relevance improvement while replacing most manual work in the iteration pipeline.
Extensive human evaluation shows that it performs effectively across relevance-related tasks, improves annotation accuracy, and enables more timely and generalizable bad-case resolution.
These results suggest a practical paradigm for industrial search relevance optimization: evolving relevance not only through stronger models but also through autonomous multi-agent frameworks.

\section{Relevance Task Definition}
\label{sec:task_definition}

\paragraph{Task.}
In international e-commerce search, relevance measures how well a product matches the user intent expressed in a query.
Given a query $q \in \mathcal{Q}$ and a candidate product (document) $d \in \mathcal{D}$, a relevance model predicts a discrete label
$y \in \mathcal{Y}$, where we use a four-level taxonomy $\mathcal{Y}=\{0,1,2,3\}$ corresponding to
\textit{Irrelevant}, \textit{Weakly Relevant}, \textit{Relevant}, and \textit{Strongly Relevant}.
We denote the model prediction as
\begin{equation}
\hat{y}=f_{\theta}(q,d),
\end{equation}
where $\theta$ are model parameters.

\paragraph{Standards, data, and model.}
A key property of industrial relevance is that the label space is operationalized by natural-language relevance standards written by domain experts.
These standards are inherently incomplete and must evolve over time due to (i) diverse user preferences across regions and languages and
(ii) the difficulty of exhaustively specifying edge cases in text.
We denote the current standards as a context variable $S \in \mathcal{S}$ and treat them as the interface that grounds annotations and model behavior.
In addition, production systems often require time-critical, high-priority directives (e.g., compliance constraints, promotions, or emergency fixes); we model them as an optional intervention context $I \in \mathcal{I}$ injected at inference time.
Accordingly, the relevance predictor can be written as a conditional model
\begin{equation}
\hat{y}=f_{\theta}(q,d \mid S,I),
\end{equation}
where $I$ is set to an empty context by default.
The training set is constructed by annotating query--product pairs under $S$.

\paragraph{Bad cases and optimization objective.}
Let $y^{*}$ be the reference label for $(q,d)$ established under the current standards (e.g., PM adjudication or high-confidence review).
We define a \textbf{bad case} as any instance where the online model prediction disagrees with the reference label:
\begin{equation}
\mathbb{I}_{\mathrm{bad}}(q,d;S,I,\theta)=\mathbf{1}\!\left[f_{\theta}(q,d \mid S,I)\neq y^{*}\right].
\end{equation}
Our primary objective is to minimize the rate of bad cases on online traffic:
\begin{equation}
\mathcal{R}_{\mathrm{bad}}(\theta;S,I)
=\mathbb{E}_{(q,d,y^{*})\sim \mathcal{P}_{\mathrm{online}}}
\left[\mathbb{I}_{\mathrm{bad}}(q,d;S,I,\theta)\right].
\end{equation}
Here, $\mathcal{P}_{\mathrm{online}}$ denotes the online traffic distribution induced by real user queries and the serving pipeline.

We reduce $\mathcal{R}_{\mathrm{bad}}$ through two complementary mechanisms.
First, we improve the supervised signal by refining standards and data, which in turn improves the model parameters $\theta$ through retraining.
Second, we apply online interventions by updating $I$ to immediately control model behavior without waiting for a full training cycle.
The remainder of this paper presents a practical framework that operationalizes these levers via multi-agent automation (Section~\ref{sec:multi_agent})
and harness engineering extensions (Section~\ref{sec:multi_agent_extensions}).

\section{Multi-Agent Framework}
\label{sec:multi_agent}

As shown in Figure~\ref{fig:framework}, we operationalize relevance iteration as a case-driven loop executed by specialized agents that jointly minimize the online bad-case rate $\mathcal{R}_{\mathrm{bad}}(\theta;S,I)$ (Section~\ref{sec:task_definition}).
Each agent corresponds to one operational lever in the bad-case reduction loop: case discovery, standard-grounded adjudication, and repair.
The loop combines an \textbf{Annotator Agent} for standard-grounded labeling (Section~\ref{subsec:annotator}), an \textbf{Optimizer Agent} for data-centric repair of model failures (Section~\ref{subsec:optimizer}), and a \textbf{User--Annotator} dialectic for autonomous bad-case discovery and standard evolution (Section~\ref{subsec:user-expert}),
executed inside an \textbf{Automated Iteration Pipeline} for continual training and deployment (Section~\ref{subsec:automated-iteration-forge}).

\subsection{Annotator: Substitute for Human Labelers}
\label{subsec:annotator}

\begin{figure}[htb]
	\centering
	\includegraphics[width=\linewidth]{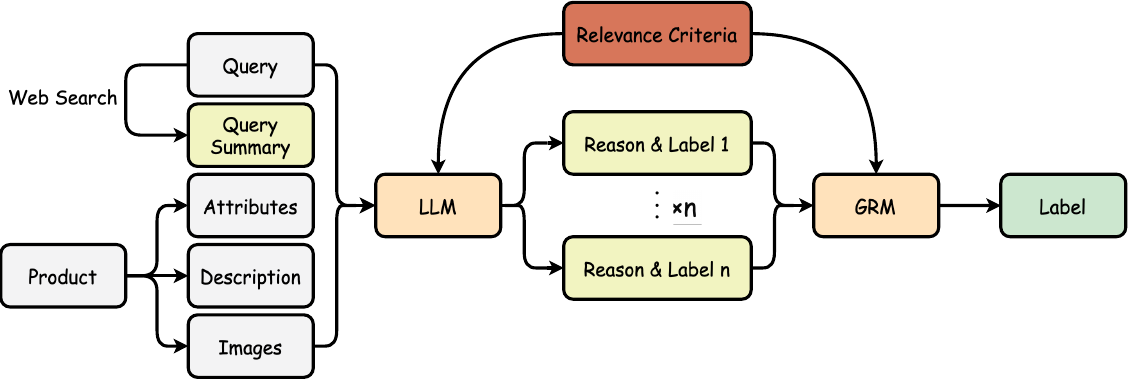}
	\caption{Annotator architecture.}
	\label{fig:annotator_arch}
\end{figure}
\noindent
Human labeling is costly, inconsistent, and slow to adapt to evolving standards.
We therefore develop an in-house, LLM-based \textbf{Annotator Agent} that produces relevance labels under the current standards $S$ (and optional directives $I$).
The design targets accurate standard-grounded labeling while minimizing hallucination and decision variance.
The pipeline (Figure~\ref{fig:annotator_arch} %in Appendix~\ref{sec:appendix_multi_agent}
) performs query grounding, uncertainty-aware candidate generation, and GRM-based selection.

\paragraph{Query grounding.}
User queries are dynamic (often long-tail or trend-driven), making pure parametric knowledge unreliable.
We retrieve open-web evidence and summarize it as a compact \textit{Query Summary} \citep{lewis2020retrieval}:
\begin{equation}
    \label{eq:query_summary}
    s_q = \text{LLM}_{\text{summary}}(q, \text{WebSearch}(q))
\end{equation}
where $q$ denotes the user query and $s_q$ represents the query summary.

\paragraph{Uncertainty-aware candidate generation.}
Given $(s_q, d, S, I)$, we sample multiple plausible judgments (with rationales) to expose uncertainty:
\begin{equation}
    \label{eq:round1_evolution}
    \mathcal{C}_1 = \{\ell_1^1, \ell_1^2, ..., \ell_1^k\} = \text{LLM}_{\text{single}}(s_q, d, S, I)
\end{equation}
where $\mathcal{C}_1$ denotes the candidate label set, $\ell$ denotes a candidate relevance label (with its rationale), and $k$ is the number of candidates.
This increases true-label coverage by 7.9\% and improves borderline-case recoverability.

\paragraph{GRM-based selection.}
We select the final label by scoring candidates with a Generative Reward Model (GRM) \citep{mahan2024generativerewardmodels}, fine-tuned from a large reranker model.
It provides stable preference-based selection under $(S,I)$ and avoids some of the instability often observed in direct LLM-as-a-judge pipelines \citep{zheng2023judging, liu2024gpt4judge}.

We compute the GRM score as a scalar preference function:
\begin{equation}
  \label{eq:grm_score}
  \text{score}(q\text{-}d,\ell)=\sigma\!\left(r_{\phi}(S,I,q\text{-}d,\ell)\right)
\end{equation}
where $q\text{-}d$ denotes the query--product pair and $r_{\phi}$ denotes a learned reward model conditioned on $(S,I)$.
The GRM-selected output is
\begin{equation}
    \label{eq:final_selection}
    \ell^* = \underset{\ell \in \mathcal{C}_1}{\text{argmax}} \; \text{score}(q\text{-}d,\ell)
\end{equation}
where $\ell^*$ denotes the selected label with the highest GRM score among the candidate set $\mathcal{C}_1$.
We train the GRM with a combined cross-entropy and pairwise preference objective:
\begin{equation}
  \label{eq:grm_pairloss}
  \mathcal{L}_{pairwise} = \log\left(1 + e^{-(\text{score}_p - \text{score}_n - \text{margin})}\right)
\end{equation}
\begin{equation}
\label{eq:grm_loss}
\mathcal{L} = \mathcal{L}_{\mathrm{CE}} + \lambda \mathcal{L}_{\mathrm{pairwise}}
\end{equation}

We further package this tool as the \textbf{Annotator Agent} and use it as the standard-grounded judge in the dialectical framework (Section~\ref{subsec:user-expert}).

\subsection{Optimizer: Automating Algorithmic Optimization}
\label{subsec:optimizer}

\begin{figure}[htb]
	\centering
	\includegraphics[width=\linewidth]{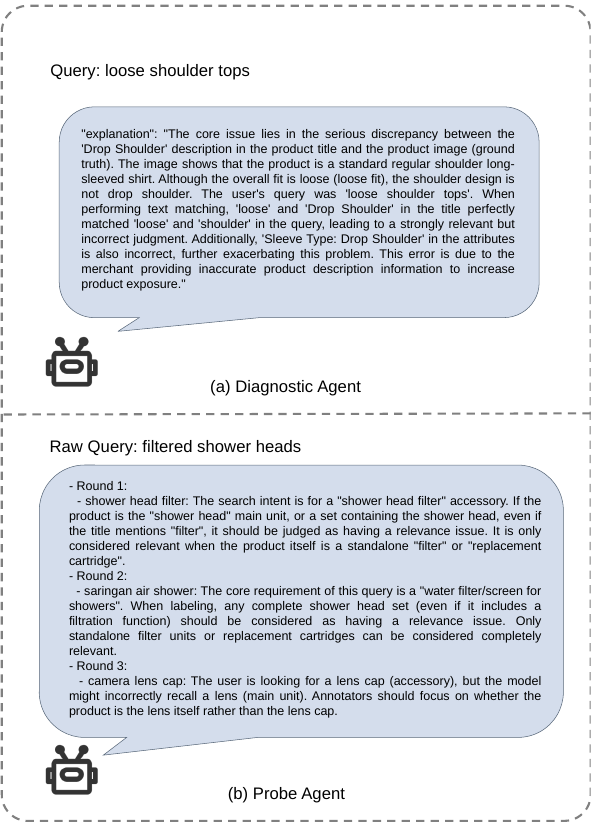}
	\caption{Running examples in the Optimizer workflow.}
	\label{fig:optimizer_flow}
\end{figure}

We propose the \textit{Optimizer}, an agentic, data-centric system for case-driven repair of e-commerce relevance models, inspired by recent calls to prioritize data quality and error analysis in ML system iteration \citep{sambasivan2021everyone, mazumder2022dataperf}.
As shown in Figure~\ref{fig:framework}(b), it decomposes the engineering loop into three agents (details in Appendix~\ref{subsec:appendix_optimizer}): a \textbf{Diagnostic Agent} (attribution and routing), a \textbf{Data Refiner Agent} (correction and augmentation), and a \textbf{Probe Agent} (pattern-level mining).
Given a reported bad case, the Optimizer first diagnoses and routes feature-side vs.\ model-side errors, then refines supervision for model-side failures, probes online traffic to generalize the pattern, and finally triggers retraining and deployment.
Formally, for $c=(q,d,y^{*})$ under $(S,I)$, let $\hat{y}=f_{\theta}(q,d\mid S,I)$ be the online prediction (Section~\ref{sec:task_definition}). Given discovered cases $\mathcal{C}$, the Diagnostic Agent performs attribution and routing:
\begin{equation}
	(\mathcal{C}_{\mathrm{feat}},\mathcal{C}_{\mathrm{model}},r)=\mathrm{Diagnose}(\mathcal{C};S,I,K),
\end{equation}
where $\mathcal{C}_{\mathrm{feat}}$ are feature-side issues, $\mathcal{C}_{\mathrm{model}}$ are model-side cases, $r$ is a structured report, and $K$ is shared memory (Section~\ref{subsec:global-memory}). The Data Refiner Agent updates supervision by correcting and augmenting data around $\mathcal{C}_{\mathrm{model}}$:
\begin{equation}
	\Delta\mathcal{D}=\mathrm{Refine}(\mathcal{C}_{\mathrm{model}},r,\mathcal{D};S,I),\quad \mathcal{D}'=\mathcal{D}\cup \Delta\mathcal{D}.
\end{equation}
The Probe Agent expands point failures into pattern-level coverage by mining additional cases through online probing:
\begin{equation}
	\Delta\mathcal{C}=\mathrm{Probe}(r;S,I,K),\quad \mathcal{C}_{\mathrm{model}}^{+}=\mathcal{C}_{\mathrm{model}}\cup \Delta\mathcal{C}.
\end{equation}
For brevity, we absorb the second refinement pass of newly mined cases into $\Delta\mathcal{D}$. The refined dataset $\mathcal{D}'$ is used to update $\theta$ via automated retraining (Section~\ref{subsec:automated-iteration-forge}), reducing $\mathcal{R}_{\mathrm{bad}}$. The Diagnostic Agent produces actionable attributions, the Data Refiner Agent corrects and augments supervision with the Annotator Agent as context, and the Probe Agent mines additional cases that match the diagnosis pattern (Figure~\ref{fig:optimizer_flow} shows examples). % in Appendix~\ref{sec:appendix_multi_agent}).

\subsection{User: Proxy for Organic User Feedback}
\label{subsec:user-expert}

While the Annotator replaces labeling and the Optimizer automates algorithmic diagnosis and repair,
industrial relevance iteration still depends on human feedback to surface difficult failures and reveal gaps in the current standards.
We therefore introduce a \textbf{User--Annotator Dialectical Mechanism} \citep{du2024improving, liang2024encouraging} that replaces organic human feedback with an agentic proxy, enabling autonomous bad-case discovery and standard evolution.
The key idea is to recreate the productive tension between \emph{experience-grounded shopping intuition} and \emph{standard-grounded adjudication}.
In this way, the mechanism not only mines high-value bad cases for model optimization but also identifies cases where the standards themselves should be revised.

\paragraph{Role definitions.}
To mirror the dialectic between user experience and formal relevance policy, we instantiate two specialized agents.

The Annotator Agent has access to the current standards $S$ and optional intervention directives $I$,
and serves as the authority on relevance criteria. Beyond pairwise judgment, it can also incorporate listwise or session-level context
when needed to maintain global consistency across a result set.
Its role is twofold: to determine whether a candidate product satisfies the current standard,
and to detect when the standard itself is insufficient to justify a consensus judgment.

The User Agent is intentionally \emph{standard-agnostic}: it does not observe $S$
and instead judges results purely from simulated shopping expectations, perceived usefulness, and outcome-oriented intuition.
This design allows it to stress-test the limits of the formal standard by surfacing failures that are obvious from experience but under-specified in policy language.

\paragraph{Dialectical bad-case discovery.}
Given a query $q$ and retrieved candidates $\{d_i\}$, the two agents first make independent judgments:
the Annotator reasons under $(S,I)$, while the User evaluates from shopping intent alone.
The agents then engage in a turn-based discussion for up to five rounds to seek consensus.
The discussion improves bad-case mining efficiency by explicitly exposing reasoning conflicts that a single-pass judge would miss.
Let $y^{\mathrm{cons}}$ denote the consensus label and $\hat{y}=f_{\theta}(q,d\mid S,I)$ the online prediction.
Formally, the discussion outcome is mapped into one of three actions.
If the consensus cannot be justified by $S$, the case is routed as a standard-evolution signal.
If the consensus is standard-consistent but disagrees with the online prediction, i.e., $y^{\mathrm{cons}}\neq \hat{y}$, the case is treated as a model error and sent to the Optimizer (Section~\ref{subsec:optimizer}) for case-driven repair.
Otherwise, the case is considered exempt: it is standard-consistent and already handled correctly online, so it is filtered out to reduce noise and conserve computation.
The detailed negotiation process is illustrated in Figure~\ref{fig:framework}(a).

\paragraph{Closing the loop.}
The User--Annotator dialectical mechanism is used in two primary scenarios.
First, it supports autonomous bad-case mining: high-precision failures are discovered continuously and routed to the Optimizer, forming a self-optimizing data flywheel for model fine-tuning and retraining.
Second, it supports interactive case handling: the same dialogue mechanism is deployed in the on-call chatbot (Section~\ref{subsec:interaction}) workflow, where engineers can inspect live cases and incorporate human adjudication when necessary. Any standard amendments or reusable precedents produced during these interactions are written into Global Memory $K$ (Section~\ref{subsec:global-memory}), allowing future judgments to benefit from prior resolutions.
Unlike traditional learning-to-rank pipelines that depend on sporadic manual feedback, this mechanism provides the missing feedback-side component needed for a fully agentic relevance ecosystem, where bad-case discovery, standard evolution, model repair, and memory accumulation can all be delegated to coordinated agents with human oversight where necessary.

\subsection{Automated Iteration Pipeline}
\label{subsec:automated-iteration-forge}

\begin{figure}[htbp]
	\centering
	\includegraphics[width=\columnwidth]{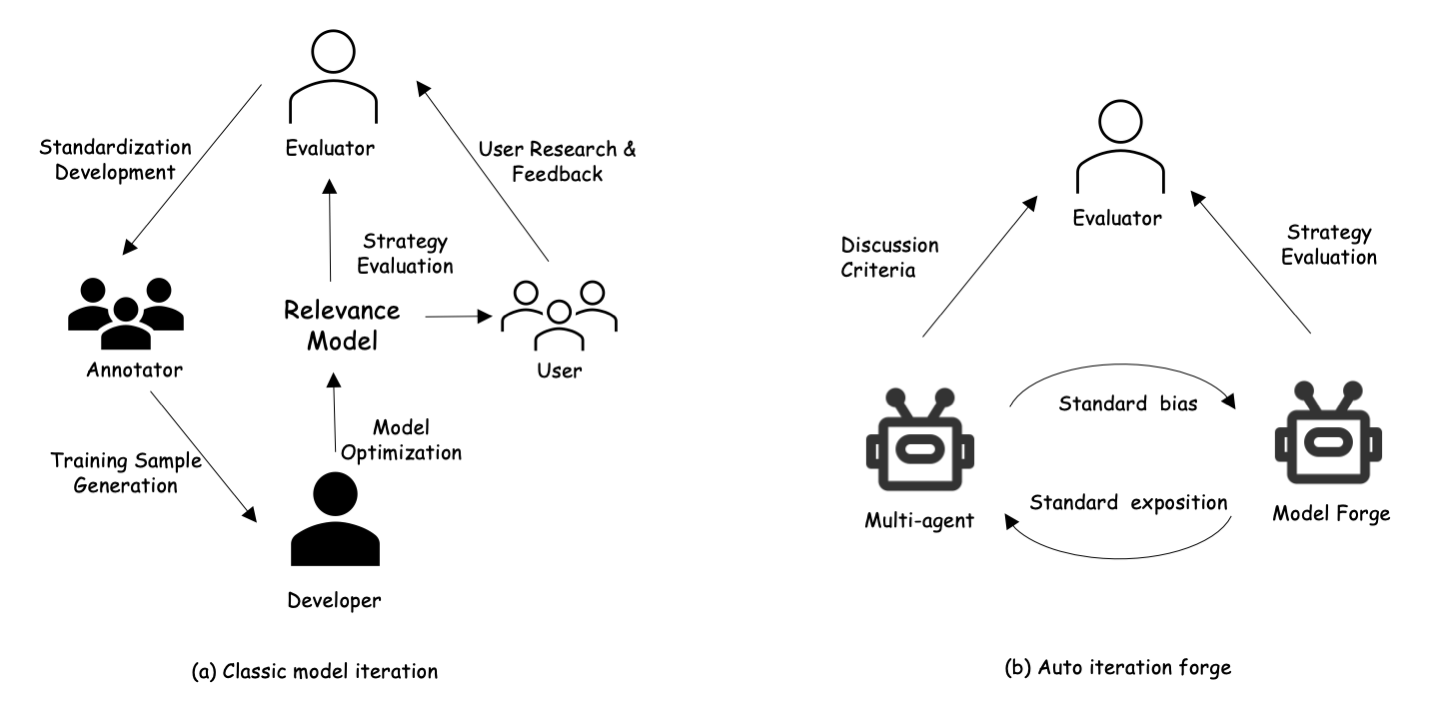}
	\caption{\textbf{Classic vs. automated iteration pipelines.} The Automated Iteration Pipeline replaces the traditional human-in-the-loop process with an agent-driven loop that continuously updates the model.}
	\label{fig:automated_forge_overview}
\end{figure}

To remove manual bottlenecks in continual updates, we build an \textit{Automated Iteration Pipeline} (Figure~\ref{fig:automated_forge_overview}) that connects the multi-agent framework with the modeling stack into a daily loop: sample online traffic, generate supervision, and train and deploy with safeguards. We maintain a cumulative corpus via
\begin{equation}
	D_t^{\mathrm{full}} = D_{t-1}^{\mathrm{full}} \cup D_t^{\mathrm{inc}}
\end{equation}
where $D_t^{\mathrm{inc}}$ is the day's incremental supervision. Scheduled jobs launch training; a checkpoint is selected with anomaly skipping and guarded by circuit breakers before promotion.
By coupling this pipeline with the multi-agent framework, the system continuously discovers, repairs, and validates bad cases with minimal human intervention, driving down $\mathcal{R}_{\mathrm{bad}}$ at a daily cadence.

While the above framework defines the decision loop, making it practical at industrial scale requires additional support from the underlying model stack and platform infrastructure.

\section{Harness Engineering Extensions}
\label{sec:multi_agent_extensions}

To make the framework practical at scale, we further build a supporting operating environment around the agents, which together instantiate a harness-engineering-style operating environment.
Concretely, these extensions operate at four layers: model unification, online controllability, cross-agent memory, and human-facing interaction.

\subsection{All-In-One Relevance Model}
\label{sec:model_approach}

\begin{figure}[htbp]
  \centering
  \includegraphics[width=\linewidth]{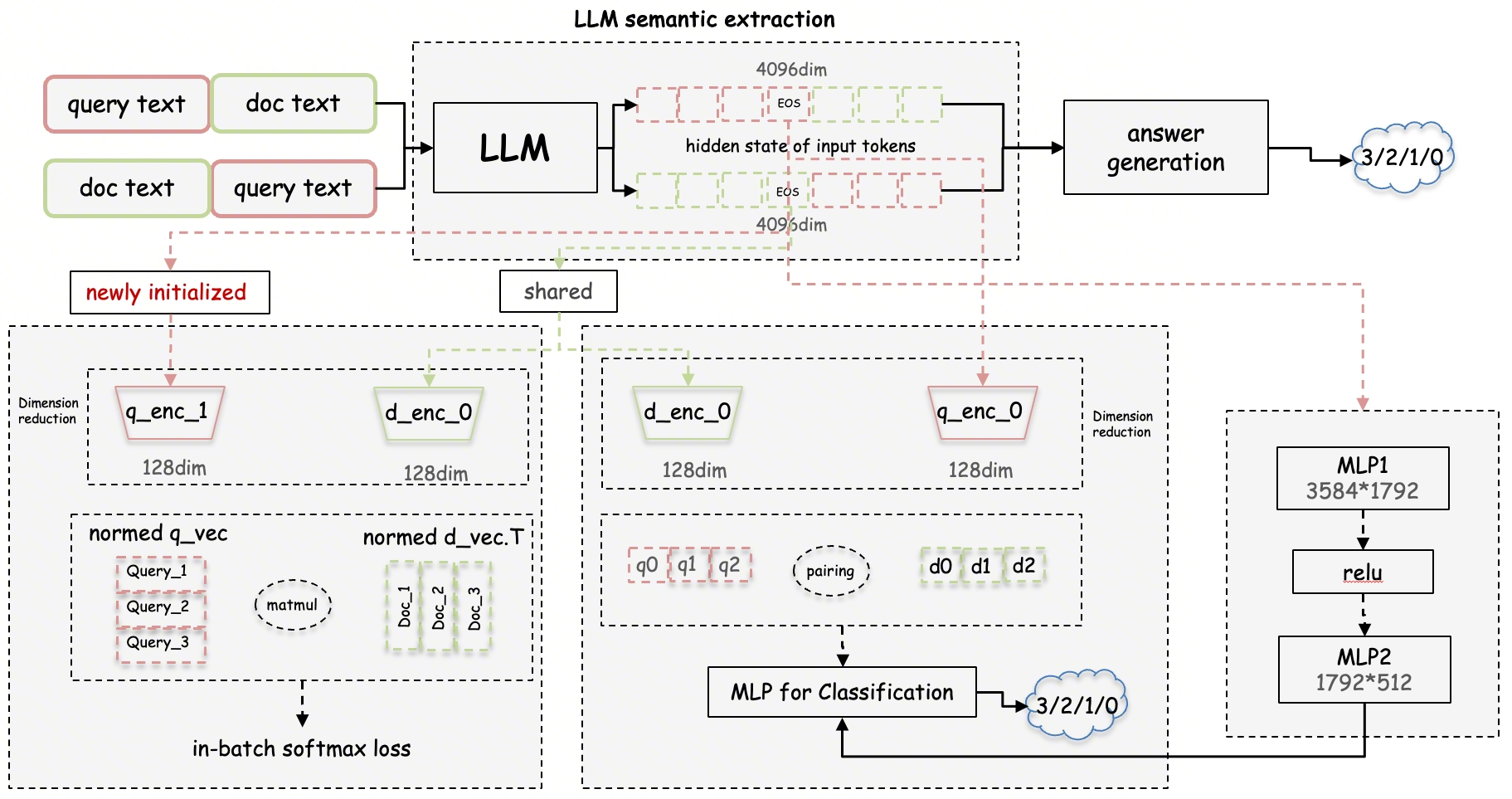}
  \caption{All-In-One model architecture.}
  \label{fig:model_arch}
\end{figure}

Our production stack follows the standard retrieval--coarse ranking--fine ranking pipeline \citep{mitra2018introduction, guo2020deep}, but instead of maintaining separate models, we adopt a unified LLM-based stack that treats these stages as different heads over a shared backbone.

The All-In-One model (Figure~\ref{fig:model_arch}) is built on a decoder-only Transformer \citep{vaswani2017attention} continue-pretrained on a large e-commerce corpus of queries, products, user interaction logs, and multilingual web data. On top of this backbone, we attach three stage-specific heads:

 \textbf{Retrieval}: a bi-encoder head that produces dense query and product representations for approximate-nearest-neighbor search over large corpora \citep{karpukhin2020dense, xiong2021approximate};
 
 \textbf{Coarse Rank}: a lightweight late-interaction head that scores thousands of candidates with moderate latency \citep{killingback2025hypencoderhypernetworksinformationretrieval};
 
 \textbf{Fine Rank}: a cross-encoder head that jointly encodes the query--product pair for high-precision relevance judgments on a small candidate set \citep{zhuang2023rankt5}.

All heads share backbone parameters and are trained jointly with a multi-task objective over retrieval, coarse ranking, and fine ranking data, which (i) reduces training and serving cost, (ii) encourages representation consistency across stages, and (iii) exposes a single relevance tool to the agents in Section~\ref{sec:multi_agent}.

\subsubsection{Query Understanding Integration}
\label{sec:query_understanding}

We further enhance the All-In-One model with \textbf{Query Understanding} (details in Appendix~\ref{sec:appendix_query_understanding}) to support query-side feature repair. This design is motivated by the Optimizer in Section~\ref{subsec:optimizer}, which identifies that some bad cases stem from defective \emph{query-side} features, such as typos, ambiguity, or missing semantic structure. To mitigate these issues, a nearline LLM parses each query into structured attributes such as category, brand, and salient properties, and the outputs are cached and injected into all stages of the model. In parallel, a query-correction augmenter mines typo--correction pairs from logs and injects corrected variants into training data. Together, these mechanisms improve robustness to noisy input and provide higher-quality query representations for both ranking and agent-based optimization. In offline evaluation, the Query Structure LLM improves F1 by 3.9\% on attribute recognition and 2.7\% on brand recognition over the production BERT baselines; when deployed with query correction augmentation, it yields a cumulative 6.91\% SBS win-rate gain online.
Document-side feature issues are handled by other teams and are beyond the scope of this paper.

\subsubsection{Resource Optimization}
\label{sec:resource_optimize}

Even with a unified model, industrial deployment must balance quality and cost. We therefore add two resource-aware mechanisms (details in Appendix~\ref{sec:appendix_resource_optimize}) on top of the All-In-One stack: a coarse--fine joint inference strategy that routes \emph{easy} high-frequency queries to a cheaper path while reserving fine ranking for long-tail or ambiguous cases, and a structured query--product relevance cache that reuses fine-grained relevance scores across queries sharing category, brand, and key attributes. Together, these mechanisms substantially reduce GPU load by 21\% while keeping online relevance metrics stable.

\subsection{Instruction-Following Capability of Relevance Model}
\label{subsec:annotator-instruct}

Large-scale promotions and sudden policy changes require ranking and retrieval to react within minutes, whereas the case-driven loop of logging failures, annotating data, and retraining models is inherently slower. To operationalize the intervention context $I$ in Section~\ref{sec:task_definition}, we turn the relevance model into an instruction-following tool that can be steered online by natural-language rules (details in Appendix~\ref{subsec:appendix_annotator-instruct}).

\begin{figure}[htb]
	\centering
	\includegraphics[width=\linewidth]{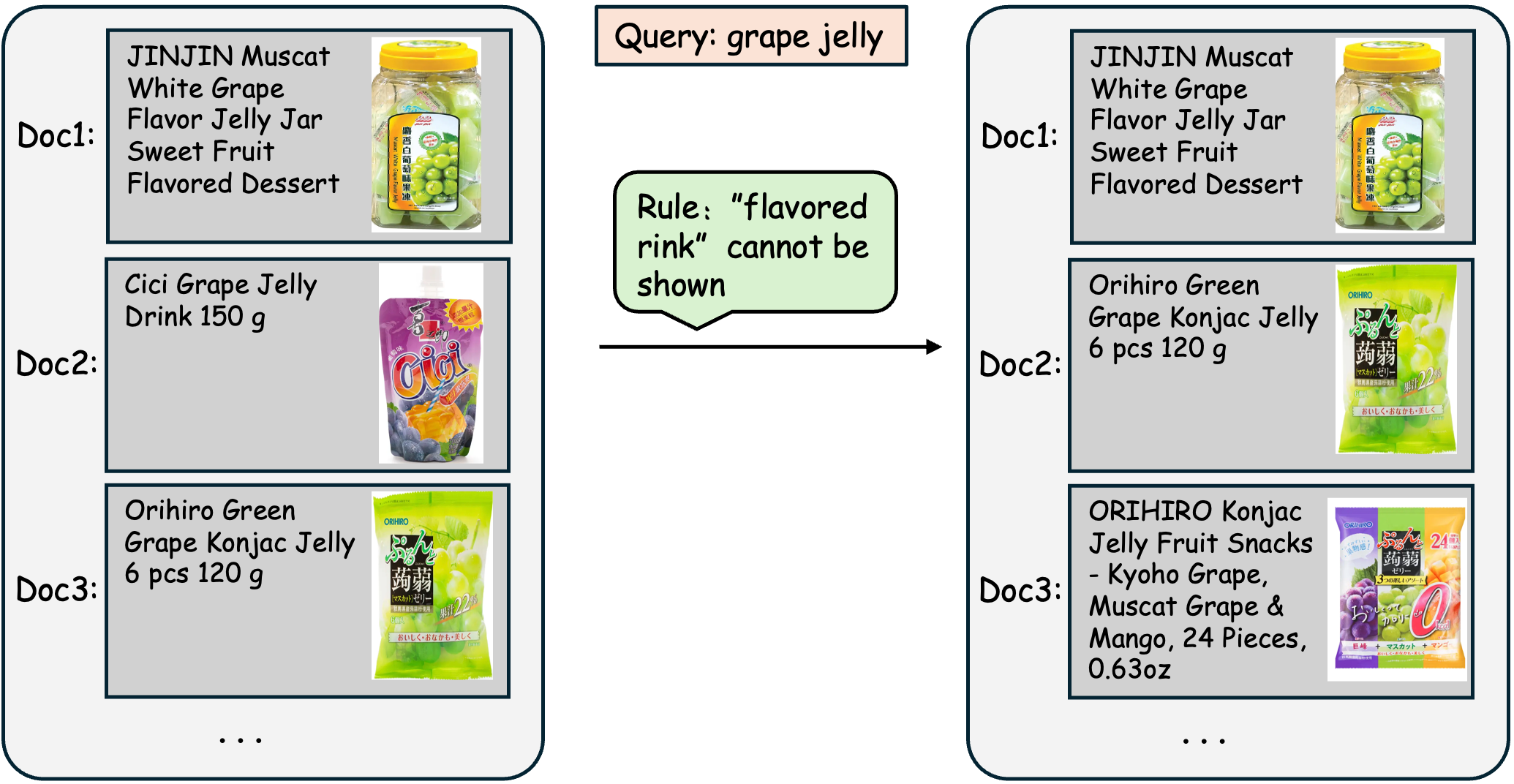}
	\caption{Examples of rules for injecting instructions into the model.}
	\label{instruct_example}
\end{figure}

We adopt an \textbf{offline instruction training + online rule injection} paradigm \citep{ouyang2022instruct, wei2022finetuned}.
Offline, we fine-tune the All-In-One relevance model so that it consumes $(q,d,S,I)$ together with an additional rule and follows a compact \texttt{Query + Product + Rule $\rightarrow$ Label + CoT} format, related to chain-of-thought prompting but optimized for low-latency classification \citep{wei2022chain}.
The model first emits the discrete relevance label and then optionally generates an explanation grounded in the rule. Online, business owners inject high-priority directives (e.g., compliance filters or emergency hotfixes) directly into the model context, as illustrated in Figure~\ref{instruct_example}, %(Appendix~\ref{subsec:appendix_annotator-instruct}),
without changing parameters $\theta$.

A key robustness challenge is avoiding \emph{rule hypersensitivity}, where the model applies any rule that appears in context. We therefore construct large Neutral and contrastive negative subsets in which the correct behavior is to ignore the injected rule and preserve the original label. Joint training on Up/Down and Neutral rules teaches the model to follow only when the rule, query, and product are strictly aligned, providing a low-latency, high-precision intervention tool that integrates naturally with the Global Memory layer in Section~\ref{subsec:global-memory}.

\subsection{Global Memory: Bridging Subjective Judgments and Objective Standards}
\label{subsec:global-memory}

A persistent challenge in search relevance is the gap between formal standards and subjective user experience,
which can cause \textit{decision oscillation} on ambiguous edge cases.
To provide shared context, we introduce a \textbf{Global Memory} system
that stores resolved precedents and curated knowledge as persistent priors,
enabling consistent decisions and continual reuse across agents and the online model \citep{park2023generative}.

The memory $K$ aggregates three complementary sources:
(i) expert-curated knowledge, including high-authority instruction rules and a small set of manually reviewed prototype cases (Sections~\ref{subsec:annotator-instruct} and \ref{subsec:user-expert});
(ii) artifacts produced by Deep Search (Section~\ref{subsec:deep-search}), such as evidence-grounded entity and attribute mappings and verified candidate associations;
and (iii) distilled traces from human--agent collaboration, periodically summarized into reusable rules and precedents (Section~\ref{subsec:interaction}).
New resolutions from these workflows are continuously distilled back into $K$ for future reuse.
At serving and deliberation time, agents retrieve and summarize relevant entries by query similarity and inject them into their decision context,
which (a) stabilizes the Optimizer's relabeling and probing decisions (Section~\ref{subsec:optimizer}),
(b) constrains the User--Annotator dialectic deliberation by aligning with settled precedents (Section~\ref{subsec:user-expert}),
(c) guides Deep Search (Section~\ref{subsec:deep-search}) when deciding whether newly retrieved candidates satisfy strong-relevance criteria,
and (d) controls behavior of the online model through providing time-critical directives.

\subsection{Deep Search}
\label{subsec:deep-search}

For long-tail queries where candidate-set misses dominate, the \textbf{Deep Search Agent} targets underestimation failures in which truly relevant products are absent from the baseline candidate generator, expanding the candidate pool so that $f_{\theta}(q,d\mid S,I)$ can operate on stronger candidates (Section~\ref{sec:task_definition}).

\begin{figure}[htbp]
	\centering
	\includegraphics[width=0.95\linewidth]{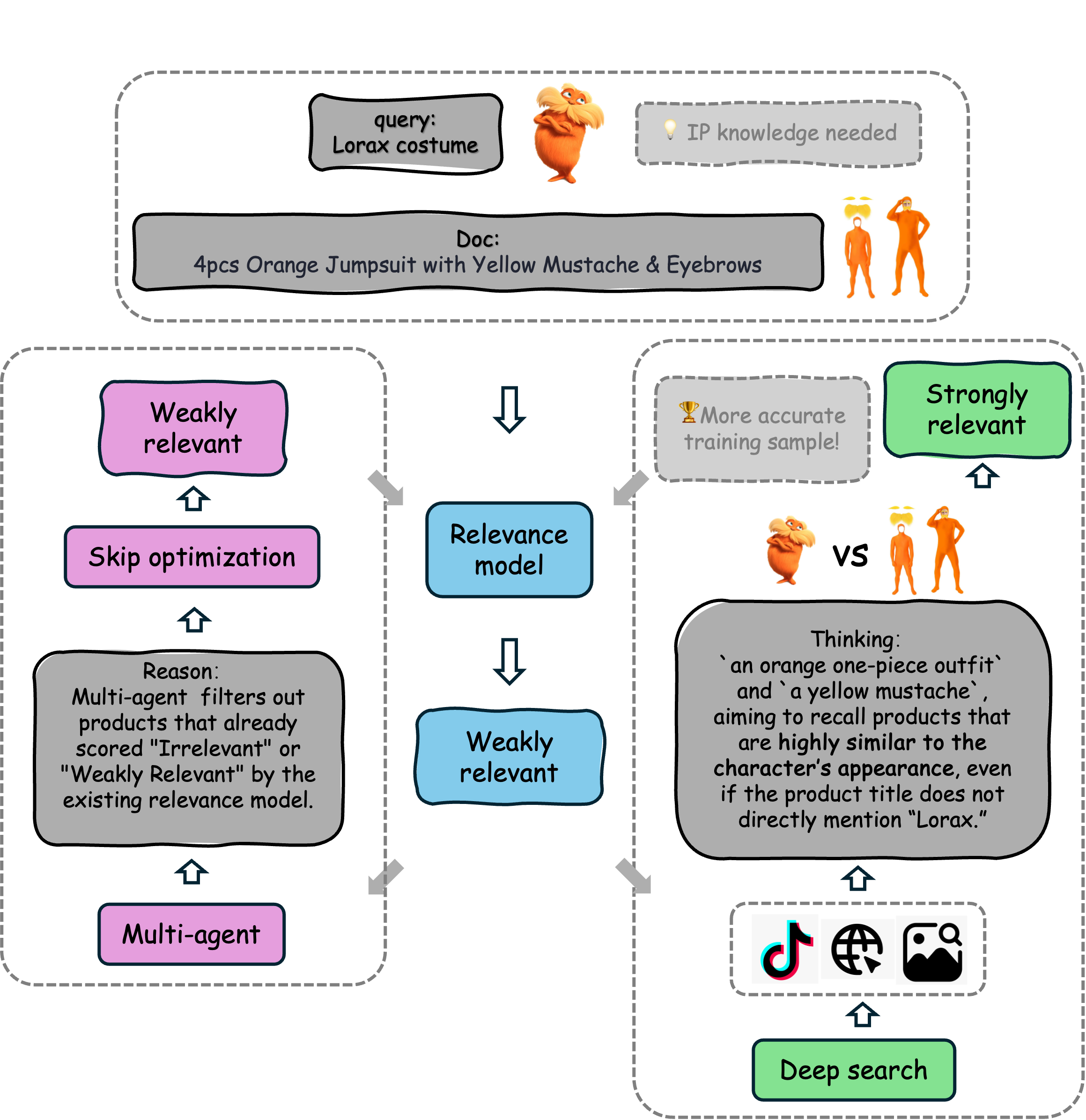}
	\caption{Deep Search improves recall for underestimated query--product pairs via an iterative retrieve--reason--act loop (illustrated on the ``Lorax costume'' case).}
	\label{fig:deep_search_workflow}
\end{figure}

Given a query $q$, the Deep Search Agent runs a stateful LLM policy $\pi$ over a tool set $\mathcal{T}$ to construct a candidate set through an iterative retrieve--reason--act loop (Figure~\ref{fig:deep_search_workflow}). Unlike one-shot query rewriting, the agent maintains an internal state $h_t$ (intent hypotheses, attempted rewrites, accumulated evidence, and candidate confidence) and updates it after each tool call:
\begin{equation}
\label{eq:deep-search-loop}
	a_t \sim \pi(\cdot \mid h_t),\quad
	h_{t+1}=\text{Update}\big(h_t,\text{Tool}(a_t)\big).
\end{equation}

The loop terminates when a step budget is reached or the top-$K$ candidates satisfy a confidence threshold.

Deep Search uses three complementary tools:
\begin{itemize}
	\item \texttt{ecom\_search}: internal vertical search over the catalog;
	\item \texttt{web\_search}: open-web retrieval for external evidence and competitor signals;
	\item \texttt{image\_search}: visual similarity retrieval within the inventory.
\end{itemize}
It supports \emph{tool chaining}, where outputs from one tool parameterize subsequent calls (e.g., web-derived image URLs $\rightarrow$ internal visual retrieval), enabling recall even when lexical overlap is weak.

To integrate Deep Search into serving, we materialize its outputs as precomputed associations
\begin{equation}
	\mathcal{M}: q \mapsto \{(d_i, w_i, \text{meta}_i)\}_{i=1}^{K},
\end{equation}
where $w_i$ is an aggregated confidence score and $\text{meta}_i$ records provenance such as tool paths. At request time the baseline candidate pool $C_{\mathrm{base}}(q)$ is augmented with Deep Search candidates,
\begin{equation}
	C_{\mathrm{aug}}(q)=C_{\mathrm{base}}(q)\cup \{d_i\}_{i=1}^{K},
\end{equation}
and the downstream predictor $f_{\theta}(q,d\mid S,I)$ is applied on $C_{\mathrm{aug}}(q)$ to mitigate underestimation on targeted query slices.

\subsection{Agent-Based Operations Chatbot}
\label{subsec:interaction}

Beyond the automated iteration loop, we deploy the multi-agent framework as a chatbot for practical case handling.
For each reported bad case, the platform orchestrates agent discussion, the relevance model, shared memory, and optional human participation
to reduce the online bad-case rate $\mathcal{R}_{\mathrm{bad}}(\theta;S,I)$ (Section~\ref{sec:task_definition}) via:

 \textbf{Updating $\theta$ (case-driven optimization)}: standard-consistent bad cases are sent to the Optimizer, and the final resolution is summarized and written into Global Memory.
 
 \textbf{Updating $I$ (online intervention)}: the bot injects time-critical directives into the online model via the instruction-following relevance tool, steering behavior without changing $\theta$ (Section~\ref{subsec:annotator-instruct}).
 
 \textbf{Human-to-agent}: human experts join the dialogue to provide adjudication and domain context, which are distilled and stored in Global Memory (Figure~\ref{fig:changyi_flow} in Appendix~\ref{subsec:agent-chatbot}).

 \textbf{Agent-to-human}: the bot emits a standard refinement suggestion for human review if the consensus cannot be justified under $S$.

\section{Experiments}
\label{sec:experiments}

We evaluate the proposed multi-agent framework and harness engineering extensions using a combination of online side-by-side (SBS) experiments and offline metrics.
Unless otherwise noted, SBS win-rate gain is measured on randomized samples of production queries with the immediately preceding strategy as the baseline.

\subsection{Multi-Agent Framework}
We first assess the end-to-end multi-agent framework on online traffic from an e-commerce marketplace.
The initial baseline model is trained on a dataset annotated via a crowdsourcing platform, and each subsequent strategy uses the immediately preceding strategy as its SBS control.

\textbf{Annotator.}
For each Annotator strategy we re-annotate the full training set, so every change corresponds to a new round of labeling and retraining.
On our offline test set, the LLM-based Annotator improves labeling precision by 2.4\% over human annotators, reduces labeling costs by 75.4\%,
and GRM-based selection further improves labeling precision by 1.3\%.

\textbf{Optimizer.}
For the Optimizer, we feed bad cases from quarterly offline evaluations into the agent.
Based on these signals, the Optimizer (i) corrects similar erroneous samples in the existing training data,
(ii) uses a research mode to search online in real time for related bad cases and construct additional training samples,
and (iii) directly generates positive and negative examples via LLM for queries with little or no supply.

\textbf{User--Annotator.}
To evaluate the User--Annotator discussion strategy, we construct a high-quality offline bad-case dataset reviewed by domain experts, sampled from top results of random queries.
We then run different agent configurations on the same query--candidate sets and compare their outputs against the reference to compute bad-case precision and recall,
as summarized in Table~\ref{tab:multi_agent_eval}.
This mirrors the intended usage of the framework: the goal is to discover high-value failures with high precision rather than to maximize overall label accuracy.
The mined bad cases are subsequently sent to the Optimizer Agent to improve the online model.

\begin{table}[t]
\centering
\small
\begin{tabular}{lcc}
\hline
\textbf{Method} & \textbf{Precision} & \textbf{Recall} \\ \hline
Human baseline & 0.9545 & 0.8289 \\ \hline
Annotator Agent (initial) & 0.7442 & 0.8421 \\
User Agent (initial) & 0.7403 & 0.7500 \\ \hline
\textbf{User--Annotator discuss} & \textbf{0.9524} & 0.7895 \\ \hline
\end{tabular}
\caption{Comparison of bad-case mining performance. Discussion between the User and Annotator agents bridges the gap between raw experience and formal standards. Each row compares the listed strategy against the same common production baseline.}
\label{tab:multi_agent_eval}
\end{table}

\begin{figure}[htbp]
	\centering
	\includegraphics[width=\columnwidth]{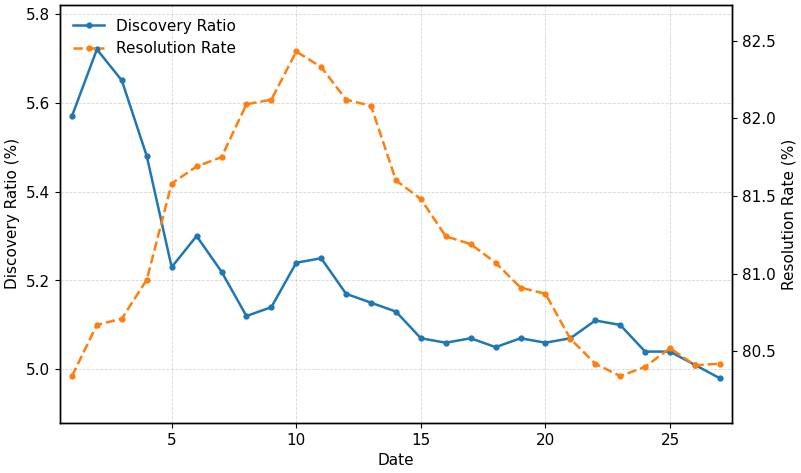}
	\caption{Trends of discovery rate and resolution rate over successive automated iterations.}
	\label{fig:rates_trend}
\end{figure}

\textbf{Automated Iteration Pipeline.}
We monitor \emph{discovery rate} (discovered failures / crawled samples) and \emph{resolution rate} (correct predictions / discovered cases).
Figure~\ref{fig:rates_trend} shows that the resolution rate exhibits an inverted-U shape as remaining cases become harder. We periodically run offline and online evaluations; Table~\ref{tab:evaluation_results} reports SBS win-rate gain across phases, reaching 4.36\% in Phase~III.

\begin{table}[htbp]
	\centering
	\begin{tabular}{lcc}
		\hline
		\textbf{Phase} & \textbf{Win-Rate Gain} \\ \hline
		Phase I & 2.71\% \\
		Phase II  & 3.1\% \\
		Phase III & 4.36\% \\ \hline
	\end{tabular}
	\caption{Relevance evaluation results across different evolution phases. Each row reports the incremental SBS win-rate gain over the immediately preceding deployed strategy.}
	\label{tab:evaluation_results}
\end{table}

\begin{table}[t]
	\centering
	\begin{tabular}{lcc}
		\textbf{Strategy} & \textbf{Win-Rate Gain} \\
		\hline
		Crowdsource baseline & - \\
		LLM Annotator & 12.38\% \\
		GRM-based & 9.78\% \\
        Agentic Optimizer  & 7.28\% \\
		User--Annotator  & 8.9\% \\
		Automated Pipeline & 10.17\% \\
	\end{tabular}
    \caption{Results of multi-agent experiments. Each row reports the incremental SBS win-rate gain over the immediately preceding deployed strategy.}
	\label{tab:model_comparison_annotator}
\end{table}

As shown in Table~\ref{tab:model_comparison_annotator}, each optimization strategy yields a positive SBS win-rate gain on the main market.
When we synchronize these strategies to a less-commonly spoken language market, we observe a cumulative SBS win-rate improvement of \textbf{13\%},
indicating that the multi-agent framework transfers robustly across markets.

\subsection{Harness Engineering Extensions}

Building on the multi-agent framework, we design and evaluate a set of system extensions that make the framework more efficient and reliable in production.

\textbf{All-In-One Relevance Model.} We conduct offline experiments on an internal e-commerce relevance dataset containing 4 million query--product pairs with multi-agent-framework annotations.
The test set is a separate in-domain expert-labeled collection of about 14k query--product pairs.
All experiments are conducted on 8$\times$ 80~GB GPUs with a global batch size of 128. Optimization follows AdamW \citep{loshchilov2019adamw}.
As shown in Table~\ref{tab:model_comparison_all_in_one}, adopting the All-In-One LLM architecture substantially increases the accuracy of the Coarse Rank stage on the offline test set, with no decrease in the Retrieval and Fine Rank stages.
\begin{table}[t]
	\centering
	\begin{tabular}{lcccc}
		\textbf{Model version} & \textbf{Precision} & \textbf{Recall} & \textbf{ACC} \\
		\hline
		Retrieval base & - & 0.858 & - \\
		Retrieval new & - & 0.858 & - \\
		\hline
		Coarse Rank base  & 0.859 & 0.881 & 0.798 \\
		Coarse Rank new  & 0.885 & 0.874 & 0.815 \\
		\hline
		Fine Rank base  & 0.922 & 0.898 & 0.857 \\
		Fine Rank new  & 0.924 & 0.897 & 0.857 \\
	\end{tabular}
	\caption{All-In-One experiment results. Baselines are separate models for the three stages.}
	\label{tab:model_comparison_all_in_one}
\end{table}

\textbf{Instruction-Following Capability of Relevance Model.}
We evaluate whether the model (i) executes applicable rules and (ii) ignores irrelevant ones.
The test set spans three scenarios---\textbf{Up} (promotion), \textbf{Down} (demotion), and \textbf{Neutral} (rule-irrelevant)---with sizes 20k/20k/100k, respectively.
Neutral is emphasized because most production traffic should not match any emergency directive.
As shown in Table~\ref{tab:instruct_follow_results}, we report overall accuracy and per-scenario accuracies ($\mathrm{ACC}_{\uparrow}$, $\mathrm{ACC}_{\downarrow}$, $\mathrm{ACC}_{\rightarrow}$).

\begin{table}[t]
\centering
\small
\setlength{\tabcolsep}{10pt}
\renewcommand{\arraystretch}{1.15}
\begin{tabular}{l|c}
\hline
\textbf{Training recipe} &
\begin{tabular}[c]{@{}l r@{}}
\textbf{ACC} & \\
\end{tabular}
\\
\hline
\textbf{Positive-only} &
\begin{tabular}[c]{@{}l r@{}}
Total & 0.5033 \\
Up $\uparrow$ & 0.8959 \\
Down $\downarrow$ & 0.9731 \\
Neutral $\rightarrow$ & 0.3848 \\
\end{tabular}
\\
\hline
\makecell[l]{\textbf{+ Contrastive}\\\ \textbf{neutral negatives}}  &
\begin{tabular}[c]{@{}l r@{}}
Total & 0.8402 \\
Up $\uparrow$ & 0.8779 \\
Down $\downarrow$ & 0.9088 \\
Neutral $\rightarrow$ & 0.8270 \\
\end{tabular}
\\
\hline
\end{tabular}
\caption{Instruction-following evaluation on Up/Down/Neutral scenarios. Introducing neutral contrastive negatives substantially improves robustness, especially on Neutral, leading to the best overall accuracy.}
\label{tab:instruct_follow_results}
\end{table}

\textbf{Global Memory.} We implement memory retrieval based on vector search. For heterogeneous memory types, we perform summaries at different levels of abstraction and then inject these summaries into the context of the multi-agent framework and the online model.

\textbf{Deep Search.}
We evaluate Deep Search on query slices where the online stack is known to underestimate relevance, including long-tail attribute combinations, scenario-based queries, and domains with sparse catalog coverage.
For each slice, we run Deep Search offline to generate candidates, send the candidate sets through the Annotator Agent for high-quality labeling, and retain only strongly relevant candidates in a nearline recall booster.

As shown in Table~\ref{tab:comparison_harness_engineering}, after applying this series of harness-engineering methods to our multi-agent framework, we further achieve significant online SBS win-rate gains.

\begin{table}[t]
	\centering
	\begin{tabular}{lcc}
		\textbf{Strategy} & \textbf{Win-Rate Gain } \\
		\hline
		All-In-One LLM & 4.56\% \\
		Instruction-Following & 3.94\% \\
		Global Memory  & 2.92\% \\
		Deep Search  & 5.84\% \\
	\end{tabular}
	\caption{Results of harness-engineering extension experiments. Each row reports the incremental SBS win-rate gain over the immediately preceding deployed strategy.}
	\label{tab:comparison_harness_engineering}
\end{table}

\section{Conclusion}

We presented a case-driven multi-agent framework for e-commerce search relevance that automates the full loop from bad-case discovery to annotation, diagnosis, repair, and deployment. Instead of treating relevance improvement solely as a modeling problem, we formulate it as a coordinated system in which specialized agents substitute key human roles in the traditional iteration pipeline.

To make this framework practical in production, we further implement harness engineering with unified retrieval and ranking, an instruction-following relevance model, the Deep Search Agent, and shared Global Memory. Together, these components enable faster annotation, more efficient bad-case resolution, and more responsive online control.

Extensive experiments and production deployment demonstrate that the framework improves annotation quality, accelerates bad-case resolution, and yields consistent online gains. Overall, our results indicate that an autonomous multi-agent framework provides a practical paradigm for industrial search relevance optimization.

\section*{Limitations}

Our work has several limitations. First, the framework is developed and evaluated in a single industrial e-commerce environment, and its effectiveness may depend on platform-specific factors such as traffic distribution, catalog structure, business rules, and annotation practices. Its generalization to other domains, markets, or regulatory settings remains to be validated.

Second, the proposed agents only approximate the human roles they replace. LLM-based agents may hallucinate, overfit to shallow cues, or produce overconfident judgments on ambiguous cases. Multi-agent discussion reduces but does not fully eliminate these risks, especially when agents share similar underlying model biases.

Third, our current framework mainly automates high-frequency operational roles such as case discovery, annotation, and repair, but does not fully replace PM- and evaluator-like functions. In addition, while agents can propose refinements to relevance standards, the governance of standard updates remains only partially addressed in the current system.

Finally, the framework incurs considerable engineering and computational cost. Multi-agent orchestration, memory maintenance, and LLM serving increase system complexity and deployment overhead. Reducing this cost while preserving effectiveness is an important direction for future work.

\section*{Ethical Considerations}

E-commerce relevance systems directly affect product exposure and user choice, making fairness, safety, and privacy central concerns. Because our framework uses LLM-based agents for annotation and decision support, it may inherit or amplify existing biases across merchants, brands, regions, or user groups. To mitigate this risk, we monitor performance across key slices, analyze systematic bad cases, and incorporate fairness-related constraints into relevance standards and online rules.

The instruction-following capability of the relevance model can help enforce compliance and platform policies, but it also creates risks when rules are overly broad or incorrectly configured. To reduce such risks, high-impact interventions should be deployed in a controlled manner, with rationales and case records retained for auditing.

\section*{Acknowledgments}

We thank our colleagues in search relevance, engineering infrastructure, product, annotation, and operations for their support in designing, deploying, and maintaining the systems described in this work, and we are grateful to the platform and tooling teams behind the internal LLM stack, agent orchestration, and monitoring infrastructure, whose contributions made large-scale experimentation and production deployment possible.

\bibliography{custom}
\appendix

\section{Related Work}
\label{sec:appendix_related_works}

Industrial search relevance has long been studied as a learning-to-rank problem, where gradient-boosted decision trees and neural scoring models are trained on large collections of human- or click-labeled data \citep{liu2009learning, mitra2018introduction, guo2020deep}.
While these systems deliver strong performance on head queries, they often rely on hand-crafted features, relatively static feedback loops, and manual case-based debugging.

Transformer-based language models have since led to dense retrievers, cross-encoders, and multi-stage retrieval-and-ranking pipelines \citep{devlin2019bert, karpukhin2020dense, khattab2020colbert, nogueira2019passage, zhuang2023rankt5}
that model query--document semantics more effectively across languages and modalities.
In parallel, instruction-tuned LLMs have been explored both as rankers and as teachers that provide labels, explanations, or synthetic data for downstream models, often combined with retrieval-augmented generation and chain-of-thought prompting to expose reasoning and better align relevance decisions with natural-language standards \citep{chung2024flan,ouyang2022instruct,lewis2020rag,wei2022cot}.

Recent work on agents and multi-agent systems views retrieval, diagnosis, and optimization as sequential decision-making problems, with specialized agents acting as annotators, critics, planners, or user proxies and invoking tools such as search engines and LLM-based models \citep{yao2023react, du2024improving, park2023generative, wu2024autogen,li2023camel,hong2024metagpt}.
Our work operationalizes these ideas in a production e-commerce setting by combining an All-In-One, instruction-following relevance LLM with Global Memory and a case-driven multi-agent framework that automates bad-case discovery, diagnosis, and repair within a single auditable platform.

\section{Full Author List}
\label{sec:appendix_author_list}

\noindent
\makebox[0.45\linewidth][l]{Lufeng Yang}
\makebox[0.45\linewidth][l]{Junfeng Hu} \\
\makebox[0.45\linewidth][l]{Feiyi Wang}
\makebox[0.45\linewidth][l]{Yonghui Huang}\\
\makebox[0.45\linewidth][l]{Wenjun Yan}
\makebox[0.45\linewidth][l]{Weidong Hou}\\
\makebox[0.45\linewidth][l]{Yuxuan Jiang}
\makebox[0.45\linewidth][l]{Dan Li}\\
\makebox[0.45\linewidth][l]{Xiaoyu Liu}
\makebox[0.45\linewidth][l]{Ya Xiao}\\
\makebox[0.45\linewidth][l]{Lingfeng Dai}
\makebox[0.45\linewidth][l]{Zhunheng Wang}\\
\makebox[0.45\linewidth][l]{Ningjia Fu}
\makebox[0.45\linewidth][l]{Guobing Wang}\\
\makebox[0.45\linewidth][l]{Mengsha Liu}
\makebox[0.45\linewidth][l]{Kaiwen Li}

\section{Multi-Agent Framework}
\label{sec:appendix_multi_agent}

\subsection{Roles of the three contextual variables in our framework}
\label{sec:appendix_task_definition}

\begin{table*}[t]
	\centering
	\small
	\setlength{\tabcolsep}{6pt}
	\renewcommand{\arraystretch}{1.15}
	\begin{tabular}{p{0.08\linewidth} p{0.22\linewidth} p{0.2\linewidth} p{0.2\linewidth} p{0.2\linewidth}}
		\hline
		\textbf{Symbol} & \textbf{Meaning} & \textbf{Typical source / owner} & \textbf{Update frequency} & \textbf{Primary usage} \\
		\hline
		$S$ & Relevance \textbf{standards}: the relatively stable natural-language policy that defines how query--product pairs should be judged & Product managers, policy owners, expert adjudication & Slow / periodic & Grounds annotation, evaluation, and model training \\
		\hline
		$I$ & Online \textbf{intervention directives}: temporary or high-priority rules injected at inference time (e.g., compliance, promotions, emergency fixes) & Business operators, policy teams, emergency operations & Fast / on demand & Steers online model behavior without retraining \\
		\hline
		$K$ & \textbf{Global Memory}: reusable precedents, distilled case resolutions, verified evidence, and summarized human--agent interaction traces & Built from agent outputs, reviewed cases, and curated knowledge sources & Continuous / incremental & Provides shared context for agents and the online model, improving consistency and reducing decision oscillation \\
		\hline
	\end{tabular}
	\caption{Roles of the three contextual variables in our framework. $S$ defines the long-horizon relevance policy, $I$ provides short-horizon behavioral control, and $K$ stores reusable precedents and evidence for cross-agent coordination.}
	\label{tab:sik_roles}
\end{table*}

To clarify the control interface of our framework, we distinguish three complementary contextual variables: relevance standards $S$, intervention directives $I$, and Global Memory $K$. They differ in semantic role, update frequency, and operational purpose. Table~\ref{tab:sik_roles} summarizes their relationship. Intuitively, $S$ defines the default policy, $I$ applies time-critical overrides, and $K$ provides reusable precedents and evidence across agents and iterations.

\subsection{Optimizer: Automating Algorithmic Optimization}
\label{subsec:appendix_optimizer}

\paragraph{Diagnostic Agent.}
The Diagnostic Agent operates in two stages. First, in \textit{pre-feature attribution}, it compares feature consistency between the model view and the evaluation view to identify over-estimated cases caused by upstream defects, such as title SEO cheating, wrong category assignment, or missing brand information. These cases are routed to $\mathcal{C}_{\mathrm{feat}}$ and intercepted before entering the model-side negative feedback loop. Second, for cases that remain after filtering, it performs \textit{model error diagnosis}: an LLM analyzes why the model over-estimated relevance, such as neglecting key attributes, confusing semantic structure, or shifting the head word, and outputs confidence-scored root-cause tags. The result is a structured bad-case report $r$ that supports downstream probing and data refinement.

\paragraph{Data Refiner Agent.}
The Data Refiner Agent transforms analytical conclusions into measurable model gains. It performs:
\begin{itemize}[leftmargin=*]
	\item Data correction: it cleans the existing training set by identifying and rectifying mislabeled or noisy samples aligned with the diagnosed root causes, preventing the model from repeatedly learning incorrect patterns.
	\item Data augmentation: it uses generalized error patterns and hard adversarial examples generated by the Probe Agent to synthesize targeted, high-quality training samples, thereby expanding the supervised distribution toward previously uncovered difficult cases.
\end{itemize}
This execution layer closes the optimization loop by turning pattern-level diagnosis into data-centric repair.

\paragraph{Probe Agent.}
The Probe Agent mimics how human experts generalize from concrete failures to reusable patterns through three probing layers.

\textit{Concept layer.} The agent first locks onto the core semantic concept of the error via controlled perturbation. Key entities or modifiers in the query or product are added, deleted, or edited to construct forward and backward probes. If the error disappears after perturbation, the case is treated as an \textit{individual case}; if the failure persists, it is treated as a \textit{universal issue} and passed to the next layer.

\textit{Market layer.} The agent then tests cross-market generalization using multilingual alignment. Core concepts are translated into multiple languages and probed in other markets to determine whether the same semantic defect recurs. This allows single-market discoveries to be replicated globally when the failure is language-invariant.

\textit{Logic layer.} Finally, the agent abstracts specific failures into broader logical traps and generates hard adversarial examples. We implement this with a bounded ReAct-style probing cycle \citep{yao2023react} of up to three rounds: error abstraction, history review, hypothesis formation, probe design, and validation. Each hypothesis is tested with 3--5 queries. Successful replication validates the hypothesis; if all returned results are strongly relevant, the hypothesis is rejected and the agent iterates with the newly learned evidence. During probing, the agent also generates enhanced queries and annotation guidelines, which can later reinforce the same error pattern in annotator training and data construction.

\section{Harness Engineering Extensions}
\label{sec:appendix_model}

\subsection{All-In-One Relevance Model}
\label{sec:appendix_model_approach}

\subsubsection{Resource Optimization}
\label{sec:appendix_resource_optimize}

\paragraph{Joint coarse--fine inference}
The key observation is that for high-frequency, low-ambiguity queries, the coarse-ranker's relevance estimates are often nearly indistinguishable from those produced by the computationally heavier fine-ranker. This motivates a \emph{small-model + large-model} serving regime: route \emph{easy} queries to a lightweight path that omits fine ranking, while reserving the fine-ranker for long-tail and semantically demanding queries. We introduce a consistency-based downgrading strategy that decides, at request time, whether the system may substitute the fine-ranker score with the coarse-ranker score. For each query, we maintain an offline-estimated consistency score $c(q)\in[0,1]$ and downgrade when $c(q) \ge \tau$.
Let $D(q)$ denote the set of logged products associated with query $q$, and let $\mathrm{bin}_c(\cdot)$ and $\mathrm{bin}_f(\cdot)$ denote discretized relevance bins predicted by the coarse and fine models, respectively. We compute
\begin{equation}
	c(q)=\frac{1}{|D(q)|}\sum_{d\in D(q)}\mathbf{1}\!\left[\mathrm{bin}_c(q,d)=\mathrm{bin}_f(q,d)\right].
\end{equation}
To ensure statistical stability, we only estimate $c(q)$ for queries with sufficient support over a fixed aggregation window.

\paragraph{Structured query--product relevance cache}
The approach constructs generalized queries (i.e., hypernym combinations) based on structured query features such as category, brand, and attribute. When the original query misses the cache, the system automatically generates hypernym combinations for secondary retrieval. The core idea stems from the category matching principle in relevance rules: if a product does not match the query's category intent, its relevance is assigned a score of 0; consequently, more complex queries that further specify brand or attribute intents within the same category will also yield a relevance score of 0. 
For example, if a product belongs to the ``soccer shoes'' category and the current query is ``nike high-top basketball shoes,'' the system checks whether a cached result for ``nike basketball shoes'' exists in the product's history; if such a cache exists and is scored 0, the current query can be inferred to also have a relevance score of 0.
This mechanism effectively extends cache coverage through hierarchical intent matching.

\paragraph{Experiment results.}
We report online SBS win-rate gain and average GPU load for two serving strategies: joint coarse--fine inference and structured caching. As shown in Table~\ref{tab:model_resource_optimization}, both strategies reduce GPU load with negligible impact on SBS win-rate gain; together they stabilize serving and enable full-traffic LLM deployment without additional GPUs.
\begin{table}[t]
	\centering
	\begin{tabular}{lccc}
		\textbf{Strategy} & \textbf{Win-Rate Gain} & \textbf{GPU Load} \\
		\hline
		\makecell[l]{Joint coarse--fine \\ inference} & -0.88\% & -16\% \\
		\hline
		\makecell[l]{Structured \\ caching}  & -0.1\% & -5\% \\
	\end{tabular}
	\caption{Results of resource optimization experiments.}
	\label{tab:model_resource_optimization}
\end{table}

\subsubsection{Integrating Query Understanding into Relevance Models}
\label{sec:appendix_query_understanding}

\paragraph{Query Structure LLM nearline framework.}

Structured query attributes and brand information are key features for e-commerce relevance tasks. To support this, we introduce the Query Structure LLM nearline framework, which provides high-quality attribute and brand features for relevance modeling.

\textbf{Training.} We randomly sample online queries and use web search results as retrieval-augmented generation (RAG) knowledge \citep{lewis2020rag} to assist GPT-based annotation. Our Query Structure LLM is a 7B Mistral model \citep{mistralai2023mistral7b}, further pre-trained on e-commerce data and fine-tuned with SFT.

\textbf{Nearline serving.} To meet the online engine's low-latency requirements, we deploy a near real-time inference framework where the LLM continuously consumes search real-time logs and writes results to the cache. Online queries are first checked against this cache; if found, the cached result is returned. Otherwise, BERT models handle real-time inference. This approach achieves a 98\% real-time cache hit rate.

\paragraph{Query correction augmentation.}
We extract query correction pairs $\langle q, q_c \rangle$ from search logs, where $q$ is the original query and $q_c$ is its corrected version. For each query that requires correction, we augment the training data by generating an additional sample in which $q$ is replaced with $q_c$, while keeping the associated product supervision unchanged. Both the original and corrected query--product pairs are included in the training set.

\paragraph{Experiment results.}
We collect 1.2 million online queries and annotate each with structured attributes and brands using a retrieval-augmented generation pipeline with GPT as the annotator.
The resulting dataset is used to perform supervised fine-tuning of our Query Structure LLM, built on a Mistral-7B backbone.
Training is conducted on 8$\times$ 80~GB GPUs with a global batch size of 128.

\begin{table}[t]
	\centering
	\begin{tabular}{lcccc}
		\textbf{Model version} & \textbf{Precision} & \textbf{Recall} & \textbf{F1}\\
		\hline
		BERT(baseline) & 0.933 & 0.846 & 0.887\\
		LLM  & 0.964 & 0.891 & 0.926\\
	\end{tabular}
	\caption{Query-attribute experiment results.}
	\label{tab:model_comparison_query_attribute}
\end{table}

\begin{table}[t]
	\centering
	\begin{tabular}{lcccc}
		\textbf{Model version} & \textbf{Precision} & \textbf{Recall} & \textbf{F1}\\
		\hline
		BERT(baseline) & 0.870 & 0.859 & 0.864\\
		LLM  & 0.876 & 0.907 & 0.891\\
	\end{tabular}
	\caption{Query-brand experiment results.}
	\label{tab:model_comparison_query_brand}
\end{table}

We compare the proposed Query Structure LLM with the online 8-layer query-attribute BERT and query-brand BERT on their respective test sets.
As shown in Table~\ref{tab:model_comparison_query_attribute} and Table~\ref{tab:model_comparison_query_brand}, our model substantially outperforms the baselines in both dimensions, yielding a \textbf{3.9\%} F1 improvement on attribute recognition and a \textbf{2.7\%} F1 improvement on brand recognition.
These gains indicate that the Query Structure LLM can supply more accurate attribute and brand features to downstream relevance models, especially on attribute- and brand-sensitive query slices.

We also conduct an online SBS evaluation using real user queries.
By switching the relevance model's query-attribute and brand features to those produced by the Query Structure LLM nearline pipeline, and incorporating query correction augmentation, we achieve a cumulative SBS win-rate improvement of \textbf{6.91\%}.

\subsection{Instruction-Following Capability of Relevance Model}
\label{subsec:appendix_annotator-instruct}

\begin{figure*}[!t]
  \centering
  \includegraphics[width=\textwidth]{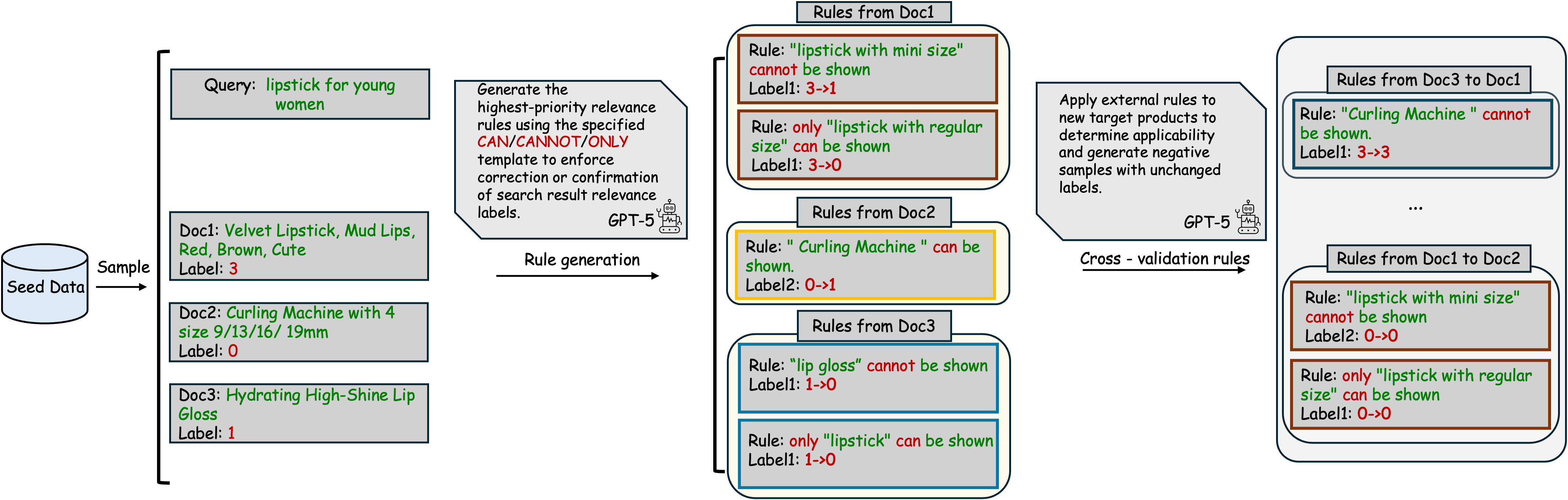}
  \caption{Workflow for generating instruction-tuning data.}
  \label{fig:workflow}
\end{figure*}

The workflow of training data generation is depicted in Figure~\ref{fig:workflow}.

\paragraph{Instruction-driven decision paradigm.}
We standardize the training data into a \texttt{Query + Product + Rule $\rightarrow$ Action + CoT} format. Here, \texttt{Query} represents the user intent, and \texttt{Product} denotes the candidate product. The \texttt{Rule} is a natural-language instruction injected into the context, specifying high-priority constraints. While the system supports various instruction forms, we focus on three core logical primitives to ground the model's reasoning capabilities:

\begin{itemize}
    \item \textbf{Inclusion (allowlist logic):} instructions that explicitly permit specific products (e.g., ``Products matching description $X$ can be shown''), often used to recall borderline relevant products.
    \item \textbf{Exclusion (blocklist logic):} instructions that strictly forbid specific attributes or brands (e.g., ``Products matching description $Y$ cannot be shown''), used for compliance and filtering.
    \item \textbf{Scoping (strict-scope logic):} instructions that define an exclusive set of valid results (e.g., ``Only products matching $Z$ can be shown''), used to lock down intent.
\end{itemize}

A representative training instance is shown below. The model first outputs the decision, then provides the justification:

\begin{center}
\fbox{\begin{minipage}{0.95\linewidth}
\small
\textbf{Input context:} \\
\textbf{Product:} Plain Summer All Seasons (Implicit Category: Women's Tanks \& Camis) \\
\textbf{Rule:} Searching for ``women's blouses'' cannot show ``women's tanks and camis'' \\
\textbf{Query:} blusas de mujer sexy (Attributes: Gender: mujer, Crowd: mujer) \\
\rule{\linewidth}{0.4pt} \\
\textbf{Model output (label + CoT):} \\
\textbf{Label:} 0 \\
\textbf{Reasoning:} The query asks for ``blusas de mujer sexy'' (women's sexy blouses). The product is explicitly a ``cami top'' and is categorized under ``Women's Tanks \& Camis.'' Per the rule, searches for women's blouses cannot show women's tanks and camis. Even though the product is sexy and for women, it is not a blouse, so the rule correctly excludes it and the relevance is none.
\end{minipage}}
\end{center}

\paragraph{Robustness via contrastive negatives.}
We construct extensive contrastive negative samples (Neutral Instructions) where the correct action is to \textit{ignore} the rule and maintain the original label.

\begin{itemize}
    \item \textbf{Object mismatch (in-query negatives):} we pair a valid rule for a specific query (e.g., ``Nike shoes cannot be shown'') with a product under the same query that does not match the rule's target (e.g., an ``Adidas'' shoe). The model must recognize the entity mismatch (Adidas $\neq$ Nike) and refuse to apply the penalty.

    \item \textbf{Scenario mismatch (cross-query negatives):} we inject rules generated for a different query context (e.g., a rule about ``sets'' from a clothing query) into an unrelated query (e.g., ``electronics''). Although the rule is syntactically valid, it is semantically out-of-scope. The model must detect this contextual conflict and suppress rule activation.
\end{itemize}

By training on these neutral scenarios, we substantially reduce false positives, ensuring the model only intervenes when the rule, query, and product are strictly aligned.

\subsection{Agent-Based Operations Chatbot}
\label{subsec:agent-chatbot}

\begin{figure*}[t]
	\centering
	\includegraphics[width=\textwidth]{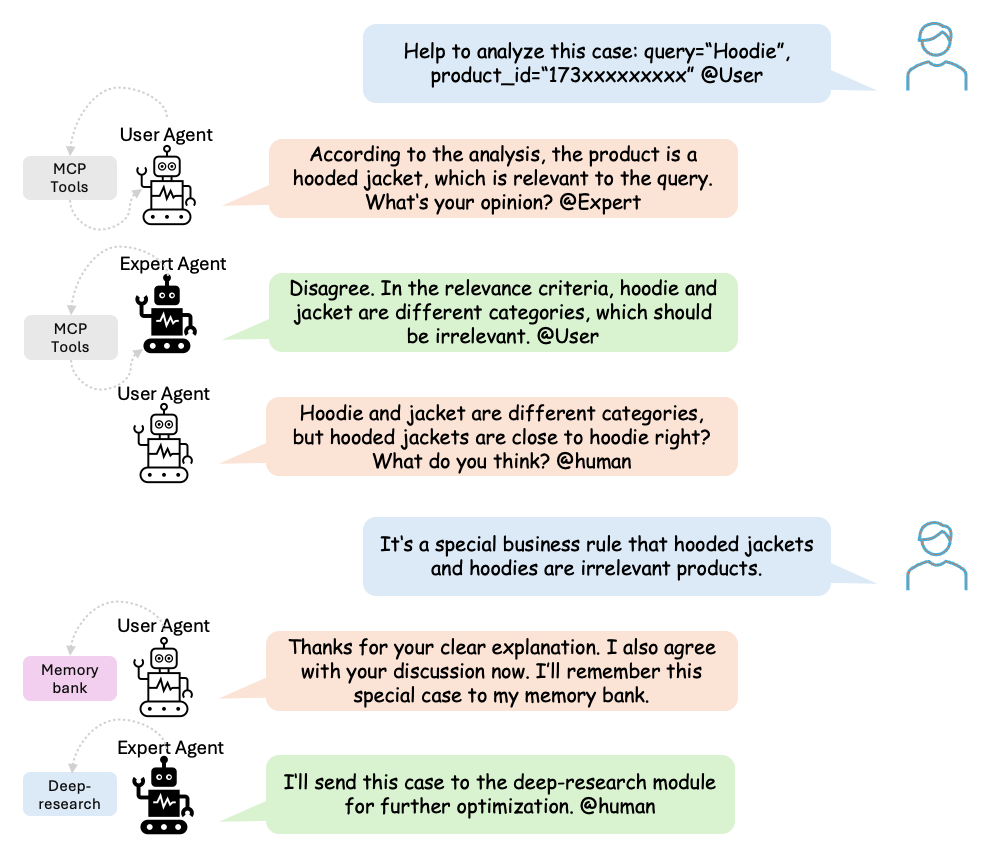}
	\caption{An example of the multi-agent conversational system. The User Agent and the Annotator (Expert) Agent discuss a difficult case with a human, invoke tool-augmented modules to resolve the case, and distill the outcome into Global Memory.}
	\label{fig:changyi_flow}
\end{figure*}

\end{document}